# On the anisotropies of magnetization and electronic transport of magnetic Weyl semimetal $Co_3Sn_2S_2$

Jianlei Shen,[1,2] Qingqi Zeng,[1,2] Shen Zhang,[1,2] Wei Tong,[3] Langsheng Ling,[3] Chuanying Xi,[3] Zhaosheng Wang,[3] Enke Liu,[1,4*] Wenhong Wang,[1,4] Guangheng Wu,[1] and Baogen Shen[1]

1. State Key Laboratory for Magnetism, Institute of Physics, Chinese Academy of Sciences, Beijing 100190, China
2. University of Chinese Academy of Sciences, Beijing 100049, China
3. Anhui Province Key Laboratory of Condensed Matter Physics at Extreme Conditions, High Magnetic Field Laboratory of the Chinese Academy of Sciences, Hefei 230031, China
4. Songshan Lake Materials Laboratory, Dongguan, Guangdong 523808, China

∗ Corresponding author.
  E-mail addresses: ekliu@iphy.ac.cn

**Abstract**

Co$_3$Sn$_2$S$_2$, a quasi-two-dimensional system with kagome lattice, has been found as a magnetic Weyl semimetal recently. In this work, the anisotropies of magnetization and transport properties of Co$_3$Sn$_2$S$_2$ were investigated. The high field measurements reveal a giant magnetocrystalline anisotropy with an out-of-plane saturation field of 0.9 kOe and an in-plane saturation field of 230 kOe at 2 K, showing a magnetocrystalline anisotropy coefficient $K_u$ up to $8.3 \times 10^5$ J m$^{-3}$, which indicates that it is extremely difficult to align the small moment of 0.29 μ$_B$/Co on the kagome lattice from *c* axis to *ab* plane. The out-of-plane angular dependences of Hall conductivity further reveal strong anisotropies in Berry curvature and ferromagnetism, and the vector directions of both are always parallel with each other. For in-plane situation, the longitudinal and transverse measurements for both I // *a* and I ⊥ *a* cases show that the transport on the kagome lattice is isotropic. These results provide essential understanding on the magnetization and transport behaviors for the magnetic Weyl semimetal Co$_3$Sn$_2$S$_2$.

Magnetic Weyl semimetals that can combine the topology and magnetism are highly desired.[1,2] Till now, the experimental realization of magnetic Weyl semimetals is still on the way. Very recently, the Shandite compound $Co_3Sn_2S_2$ was predicted as a magnetic Weyl semimetal, showing an intrinsic giant anomalous Hall conductivity and anomalous Hall angle owing to the topologically enhanced Berry curvature.[3-5] Both angle-resolved photoemission spectroscopy (ARPES) and scanning tunneling spectroscopy (STM) were performed to detect the topological state and surface Fermi arcs, which spectroscopically confirmed the prediction of magnetic Weyl semimetal phase in this system.[6,7] Nowadays, the anomalous Nernst effect,[8] negative orbital magnetism,[9] exchange bias,[10] electronic correlations,[11] spin-transfer torque,[12,13] even topological catalysis[14] have been found subsequently. Moreover, due to its long-range out-of-plane ferromagnetic order and topological band structure, this semimetal becomes an ideal candidate for developing a quantum anomalous Hall state[15] and an excellent platform for comprehensive studies on topological electronic behaviours.

The ternary Shandite compounds with the general formula $T_3M_2X_2$, where $T$ = Ni, Co, Rh or Pd; $M$ = Sn, In, or Pb; and $X$ = S or Se, have been investigated extensively in recent decades.[16-22] Among of them, $Co_3Sn_2S_2$ is the only compound showing ferromagnetism. The crystal structure of $Co_3Sn_2S_2$ (space group R-3m) possesses a quasi-2D Co-Sn kagome layers sandwiched between S atoms, and stacks in ABC fashion along the $c$-axis.[23,24] The magnetic moment of Co atoms are fixed on kagome layer in $ab$-plane and along the $c$-axis, as shown in Fig. 1(a). Although the basic properties including the magnetic anisotropy in this system have been addressed,[25] the in-depth studies on anisotropies of both magnetization and transport behaviors remain rare. In this work, we systematically investigated the anisotropies of magnetic and transport properties of $Co_3Sn_2S_2$ single crystal. The experimental results show that the magnetization presents a giant magnetocrystalline anisotropy, and angular (θ) dependence of Hall conductivity with different magnetic fields indicates strong magnetization and Berry curvature anisotropies, and the vector directions of both are always parallel with each other. Meanwhile the electronic transport on the kagome lattice is isotropic.

The single crystals of Co$_3$Sn$_2$S$_2$ can be grown by slowly cooling the melts with congruent composition (Co : Sn : S = 3 : 2 : 2).[3] The room temperature X-ray diffraction (XRD) with Cu-$K\alpha$ radiation shows only the (000*l*) Bragg peaks, which indicates that the measured plane surface of the crystal is *ab*-plane. The inset shows a naturally dissociated flake single crystal with a size of about 0.5×3×5 mm$^3$, as shown in Fig. 1(b). Magnetic properties were measured by superconducting quantum interference device (SQUID) magnetometer and the Cell-5 Magnet of the High Magnetic Field Laboratory, Chinese Academy of Sciences (CHMFL). Electrical transport was measured by the physical property measurement system (PPMS).

Figures 2(a) and (b) show the temperature dependence of magnetization with ZFC and FCC for *B* // *c* and *B* // *ab* at *B* = 200 Oe (left ordinate). A sharp magnetic transition can be observed near $T_\text{C} \approx 175$ K. Below $T_\text{C} \approx 175$ K, The magnetization for *B* // *c* is about one-hundred times larger than that of *B* // *ab*, which indicates the strong magnetocrystalline anisotropy in Co$_3$Sn$_2$S$_2$. The $1/\chi$ (T) curves from 200 to 300 K can be fitted by a modified Curie–Weiss law $\chi (T) = C/(T-\theta_\text{p})$, where *C* is the Curie constant and $\theta_\text{p}$ is the Weiss temperature. The fitting over the temperature ranging from 200 to 300 K are shown as solid yellow curves in Figs. 2(a) and (b) (right ordinate). The $m_{\textit{eff}}$ is calculated as 1.05 $\mu_\text{B}$/Co for *B* // *c* and 1.14 $\mu_\text{B}$/Co for *B* // *ab*, which are consistent with reported values.[26] The positive $\theta_\text{p}$ for two directions indicates ferromagnetic interaction in Co$_3$Sn$_2$S$_2$.

Figures 2(c) shows the field dependence of magnetization at 2 K for both *B* // *c* and *B* // *ab*. For *B* // *c*, the magnetization quickly reaches the saturation at about 0.9 kOe and the saturation magnetization $M_\text{s}$ is 10 emu/g, while for *B* // *ab* the magnetization exhibits a saturation until 230 kOe, which is about three orders of magnitude higher than that of *B* // *c*. These results further indicate the quite strong magnetocrystalline anisotropy in Co$_3$Sn$_2$S$_2$. Furthermore, the magnetocrystalline anisotropy coefficient ($K_\text{u}$) can be obtained by $K_\text{u} = \mu_0 H_\text{k} M_\text{s}/2$, where $\mu_0$ is the vacuum permeability and $H_\text{k}$ is the anisotropy field defined as the critical field above which the difference in magnetization between the two magnetic field directions (*B* // *ab* and *B* // *c*) becomes smaller than 2%.[27] The value of $K_\text{u}$ is about 8.3×10$^5$ J m$^{-3}$ at 2 K, which is larger than those of some magnetic quasi-two-dimensional materials, such as $Fe_3Sn_2$,[27] $Fe_{3-x}GeTe_2$,[28] $CrBr_3$ and $CrI_3$,[29] as shown in Table 1. In order to observe the magnetocrystalline anisotropy more clearly, we measured the angular dependent magnetization from out-of-plane to in-plane at 2 K in $B$ = 500 Oe, as shown in Fig. 2(d).

In order to further characterize the magnetocrystalline anisotropy, the anisotropic magnetocaloric effect was measured (also see supplementary materials Fig. S1). A set of magnetic isotherms M($H$) curves of $B$ // $c$ and $B$ // $ab$ were measured from 160 to 190 K. The M($H$) curves at 175 K are given in Fig. 2(e) for clarity. Based on these measurements, the magnetic entropy changes (-$\Delta S_m$) for $B$ // $c$ and $B$ // $ab$ are calculated using Maxwell relation. We show -$\Delta S_m$ at 50 kOe, as shown in Fig. 2(f). A maximum -$\Delta S_m$ value of 1.13 J kg$^{-1}$ K$^{-1}$ is obtained for $B$ // $c$ at 175 K, which is consistent with the previous reports.[30,31] For $B$ // $ab$, it is only 0.38 J kg$^{-1}$ K$^{-1}$, which indicates that there is a strong anisotropy in magnetocaloric effect. The anisotropic magnetocaloric effect further demonstrates the large magnetocrystalline anisotropy in $Co_3Sn_2S_2$.

In order to probe the relation of anisotropies between the Berry curvature and ferromagnetism, the angular ($\theta$) dependences of Hall conductivity $\sigma_{yx}$ with different magnetic fields rotating around $y$-axis in $ac$-plane have been measured and shown in Fig. 3(a). The configuration of current and magnetic field is shown in Fig. 3(b). Wang et al. reported $\rho_{yx}$ changed by only 1% when the magnetic field is tilted away from the $c$-axis up to ±30°.[5] In our work, the $\sigma_{yx}$ remains unchanged in a wide angle range indeed. In the case of 10 kOe, the stable range is as wide as more than 170°.

To understand this robust behaviour against the rotation of magnetic field, we extracted the in-plane ($B_a$) and out-of-plane ($B_c$) components of the applied magnetic field, as well as $\sigma_{yx}$ in different magnetic field at $\theta$ = 5° (we take $\theta$ = 5° as an example for the clarity), as shown in Figs. 3(c) and (d). It can be clearly seen that $\sigma_{yx}$ increases first and then decreases, reaching its maximum at 10 kOe. For $B$ = 10 kOe, $B_a$ and $B_c$ at $\theta = 5°$ are 9.96 kOe and 0.87 kOe, respectively. At this moment, $B_a$ is much lower than the in-plane saturation field 230 kOe, while $B_c$ is basically equivalent to out-of-plane saturation field. So the magnetic moment is basically along the $c$-axis and saturated. The maximum value of $\sigma_{yx}$ at $\theta = 5°$ is ~ 1170 $\Omega^{-1}$ cm$^{-1}$, which is consistent with the result of calculation based on the Berry curvature along $c$ axis.[3] For $B < 10$ kOe, the out-of-plane field cannot make the magnetic moment saturated and the single domain cannot form, so that $\sigma_{yx}$ is less than that of 10 kOe. Although $B_c$ is higher than out-of-plane saturation field, the increase of $B_a$ causes the magnetic moment to deviate away from the $c$-axis and the decrease of $\sigma_{yx}$ at the same time. The value of $\sigma_{yx}$ depends on the strength of Berry curvature along $c$ axis.[3] The decrease of $\sigma_{yx}$ indicates that the direction of Berry curvature tilts from the $c$ axis and follows the direction of magnetic moment. All these results suggest that Berry curvature has strong anisotropy as magnetic moment and the vector directions of both are keep the same and locked together. In addition, the $\varphi_M$ between magnetic moment and easy $c$-axis can be approximately estimated as $\varphi_M = B_a \times 90°/230$ kOe $= B\cos\theta(90°/230$ kOe) during the field rotation, accordingly, all the magnetic moments are aligned completely in $ab$ plane for $B = 230$ kOe applied in $ab$ plane, as shown in Fig. 3(d).

Figure 4(a) shows the temperature dependences of the longitudinal resistivity $\rho_{xx}$ in zero field and 90 kOe for both I // $a$ and I $\perp$ $a$. The configuration of current and magnetic field is shown in the inset in Fig. 4(a). It can be clearly seen that $\rho_{xx}$ completely coincide in the whole temperature range for two orientations at zero field and 90 kOe, respectively, which indicates that it is isotropic in transport for two orientations. The field-dependence of magnetoresistance also shows the consistent behavior between two orientations, as shown in Fig. 4(b).

Figure 4(c) shows the dependences of Hall resistivity $\rho_{yx}$ on temperature for I // $a$ and I $\perp$ $a$ under $B = 1$ kOe. With the decrease of temperature, the largest $\rho_{yx}$ for both orientations appear at 150 K with the same value. The dependence of the Hall resistivity $\rho_{yx}$ on magnetic field at 2 and 150 K were measured and shown in Figs. 4(d) and (e), respectively. The $\rho_{yx}$ of both orientations is consistent with each other for 2 and 150 K. The notable nonlinear field dependence of the $\rho_{yx}$ can be clearly observed and further indicates the coexistence of hole and electron carriers at 2 K for both orientations, while the single hole carriers are observed at 150 K for both orientations, which are in good agreement with previous report.[3] The $\rho_{yx}^A$ at 2 and 150 K are then obtained at the zero field. They are 0.98 and 44.6 μΩ cm for I // a, and 1.08 and 45.6 μΩ cm I ⊥ a. The dependence of $\sigma_{yx}$ on magnetic field at 2 and 150 K for both orientations are shown in Fig. 4(f). The $\sigma_{yx}^A$ at 2 K reaches up 1230 and 1300 Ω$^{-1}$ cm$^{-1}$ for I // a and I ⊥ a, respectively, which are attributed to the large Berry curvature induced by Weyl nodes and gapped nodal line near Fermi energy. In this system, there are three pairs of Weyl nodes and they are all in the same energy position, only 60 meV above the Fermi level.[3,6,7] This produces the coexistence of the Fermi surfaces of holes and electrons. Moreover, the maximum value of the anomalous Hall angles ($\sigma_{yx}^A / \sigma_{xx}$) are 18.5% and 19% at 150 K, respectively, as shown Table 2. It is evidenced that both anomalous Hall conductivity and anomalous Hall angle are isotropic on the kagome lattice.

According to band structure calculations and ARPES confirmation,[3,6] the relatively small Fermi surfaces of holes and electrons indicate the coexistence of holes and electrons, which are consistent with the notable nonlinear field dependence of the $\rho_{yx}$. Furthermore, we extracted the carrier densities and mobilities at 2 K using the semiclassical two-band model (see supplementary materials Fig. S2), as shown in Table 2. The carrier densities and mobilities for I // a and I ⊥ a are basically the same to each other, which indicates that the transport is isotropic on the kagome lattice in this magnetic Weyl semimetal. These results further reveal a near compensation of charge carriers of electrons and holes. It is considered that a $B^2$-dependence of MR described by power law MR = $AB^\alpha$ is the signature of a compensated metal.[32] For in-plane magnetic field cases,[3,25] $\alpha$ can be close to 2 in $Co_3Sn_2S_2$. However, in the case of I // $a$ and I $\perp$ $a$ under $B$ // $c$ (out-of-plane field), the $\alpha$ is only about 1.6, as shown in Fig. 4(b). One possible reason for this deviation may be the negative MR due to the spin-dependent scattering in ferromagnetic materials, which will balance out a part of the positive MR, resulting in a reduced exponent from 2. Owing to the strong magnetocrystalline anisotropy, in contrast, this negative MR effect shows weak influence on the total MR when field is applied in-plane. Therefore, a near compensation behavior can be expected in $Co_3Sn_2S_2$.

In conclusion, we investigated the anisotropies of magnetic and transport properties of magnetic Weyl semimetal $Co_3Sn_2S_2$. This semimetal exhibits strong magnetocrystalline anisotropy with out-of-plane saturation field of 0.9 kOe and in-plane saturation field of 230 kOe, and the value of $K_u$ up to $8.3 \times 10^5$ J m$^{-3}$. In addition, angular ($\theta$) dependence of Hall conductivity $\sigma_{yx}$ with different magnetic field indicates strong magnetic and Berry curvature anisotropies and the vector directions of both are locked together. For in-plane situation, the longitudinal and transverse measurements for both I // $a$ and I $\perp$ $a$ cases show that the transport on the kagome lattice is isotropic. These results provide essential understanding on the magnetization and transport behaviors for the magnetic Weyl semimetal $Co_3Sn_2S_2$, which will benefit the potential applications of the magnetic topological quantum materials.

See supplementary material for magnetization curves and magnetic entropy change measured from 160 to 190 K for $B$ // $c$ and $B$ // $ab$, respectively (Fig. S1). The fitting of longitudinal and transverse conductivities at 2 K based on two-band model to obtain carrier densities and mobilities of I // $a$ and I $\perp$ $a$ (Fig. S2 and Table S1) .

This work was supported by National Natural Science Foundation of China (Nos. 51722106 and 11974394), Beijing Natural Science Foundation (No. Z19J00024), Users with Excellence Program of Hefei Science Center CAS (No. 2019HSC-UE009), and Fujian Institute of Innovation, Chinese Academy of Sciences. A portion of this work was performed on the Steady High Magnetic Field Facilities, High Magnetic Field Laboratory, Chinese Academy of Sciences.

**References**

[1] B. Yan and C. Felser, Annu. Rev. Condens. Matter Phys. **8**, 337 (2017)

[2] Z. P. Guo, P. C. Lu, T. Chen, J. F.Wu, J. Sun, and D. Y. Xing, Sci. China Phys. Mech. Astron. **61**, 038211 (2018).

[3] E. Liu, Y. Sun, N. Kumar, L. Muchler, A. Sun, L. Jiao, S. Y. Yang, D. Liu, A. Liang, Q. Xu, J. Kroder, V. Suss, H. Borrmann, C. Shekhar, Z. Wang, C. Xi, W. Wang, W. Schnelle, S. Wirth, Y. Chen, S. T. B. Goennenwein, and C. Felser, Nat. Phys. **14**, 1125 (2018).

[4] Q. N. Xu, E. K. Liu, W. J. Shi, L. Muechler, J. Gayles, C. Felser, and Y. Sun, Phys. Rev. B **97**, 235416 (2018).

[5] Q. Wang, Y. Xu, R. Lou, Z. Liu, M. Li, Y. Huang, D. Shen, H. Weng, S. Wang, and H. Lei, Nat. Commun. **9**, 3681 (2018).

[6] D. F. Liu, A. J. Liang, E. K. Liu, Q. N. Xu, Y. W. Li, C. Chen, D, Pei, W. J. Shi, S. K. Mo, P. Dudin, T. Kim, C.Cacho, G. Li, Y. Sun, L. X. Yang, Z. K. Liu, S. Parkin, C. Felser, and Y. L. Chen, Science 365, 1282-1285 (2019).

[7] N. Morali, R. Batabyal, P. K. Nag, E. K. Liu, Q. N. Xu, Y. Sun, B. H. Yan, C. Felser, N. Avraham, and H. Beidenkopf, Science 365, 1286-1291 (2019).

[8] S. N. Guin, P. Vir, Y. Zhang, N. Kumar, S. J. Watzman, C. Fu, E. Liu, K. Manna, W. Schnelle, J. Gooth, C. Shekhar, Y. Sun, and C. Felser, Adv. Mater. **31**, e1806622 (2019).

[9] J. X. Yin, S. T. S. Zhang, G. Q. Chang, Q. Wang, S. S. Tsirkin, Z. Guguchia, B. Lian, H. B. Zhou, K. Jiang, I. Belopolski, N. Shumiya, D. Multer, M. Litskevich, T. A. Cochran, H. Lin, Z. Q. Wang, T. Neupert, S. Jia, H. C. Lei, and M. Z. Hasan, Nat. Phys. **15**, 443 (2019).

[10] E. Lachman, N. Maksimovic, R. Kealhofer, S. Haley, R. McDonald, and J. G. Analytis, arXiv:1907.06651(2019).

[11] Y. S. Xu, J. Z. Zhao, C. J. Yi, Q. Wang, Q. W. Yin, Y. L, Wang, X. L. Hu, L. Y. Wang, E. K. Liu, G. Xu, L. Lu, A. A. Soluyanov, H. C. Lei, Y. G. Shi, J. L. Luo, and Z. G. Chen, arXiv:1908.04561 (2019).

[12]Koji Kobayashi, Yuya Ominato, and Kentaro Nomura, J. Phys. Soc. Jap. **87**, 073707 (2018).

[13]D. Kurebayashi and K. Nomura, Sci. Rep. **9**, 5365 (2019).

[14]G. W. Li, Q. N. Xu, W. J. Shi, C.G. Fu, L. Jiao, M. E. Kamminga, M. Q. Yu, H. Tüysüz, N. Kumar, V. Süß, R. Saha, A. K. Srivastava, S. Wirth, G. Auffermann, J. Gooth, S. Parkin, Y. Sun, E. K. Liu, and C. Felser, Sci. Adv. **5**, eaaw9867 (2019).

[15]L. Muechler, E. K.Liu, Q. N. Xu, C. Felser. and Y. Sun, arXiv:1712.08115 (2018).

[16]R. Weihrich, A. C. Stückl, M. Zabel, and W. Schnelle, Z. Anorg. Allg. Chem. **630**, 1767 (2004).

[17]R. Weihrich, and I. Anusca, M. Zabel, Z. Anorg. Allg. Chem. **631**, 1463 (2005).

[18]R. Weihrich, and I. Anusca, Z. Anorg. Allg. Chem. **632**, 1531 (2006).

[19]R. Weihrich, S. F. Matar, V. Eyert, F. Rau, M. Zabel, M. Andratschke, I. Anusca, and T. Bernert, Prog. Solid State Chem. **35**, 309 (2007).

[20]M. Fujioka, T. Shibuya, J. Nakai, K. Yoshiyasu, Y. Sakai, Y. Takano, Y. Kamihara, and M. Matoba, Solid State Commun. **199**, 56 (2014).

[21]J. Corps, P. Vaqueiro, and A. V. Powell, J. Mater. Chem. A **1**, 6553 (2013).

[22]P. Vaqueiro and G. G. Sobany, Solid State Sciences **11**, 513 (2009).

[23]M. A. Kassem, Y. Tabata, T. Waki, and H. Nakamura, J. Crystal Growth **426**, 208 (2015).

[24]M. A. Kassem, Y. Tabata, T. Waki, and H. Nakamura, J. Solid State Chem. **233**, 8 (2016).

[25]W. Schnelle, A. Leithe-Jasper, H. Rosner, F. M. Schappacher, R. Pöttgen, F. Pielnhofer, and R. Weihrich, Phys. Rev. B **88**, 144404 (2013).

[26]T. Kubodera, H. Okabe, Y. Kamihara, and M. Matoba, Phys. B: Conden. Matter **378-380**, 1142 (2006).

[27]Z. Hou, W. Ren, B. Ding, G. Xu, Y. Wang, B. Yang, Q. Zhang, Y. Zhang, E. Liu, F. Xu, W. Wang, G. Wu, X. Zhang, B. Shen, and Z. Zhang, Adv. Mater. **29**, 1701144 (2017).

[28]Y. Liu, J. Li, J. Tao, Y. M. Zhu, and C. Petrovic, arXiv:1904.03873 (2019).

[29]N. Richter, D. Weber, F. Martin, N. Singh, U. Schwingenschlögl, B. V. Lotsch, and M. Kläui, Phys. Rev. Mater. **2**, 024004 (2018).

[30]Q. Shi, X. Zhang, E. Yang, J. Yan, X. Y. Yu, C. Sun, S. Li, and Z. W. Chen, Results in Phys. **11**, 1004 (2018).

[31]A. Ali, S., and Y. Singh, J. Appl. Phys. **126**, 155107 (2019).

[32]E. Fawcett and W. A. Reed, Phys. Rev. **131**, 2463 (1963).

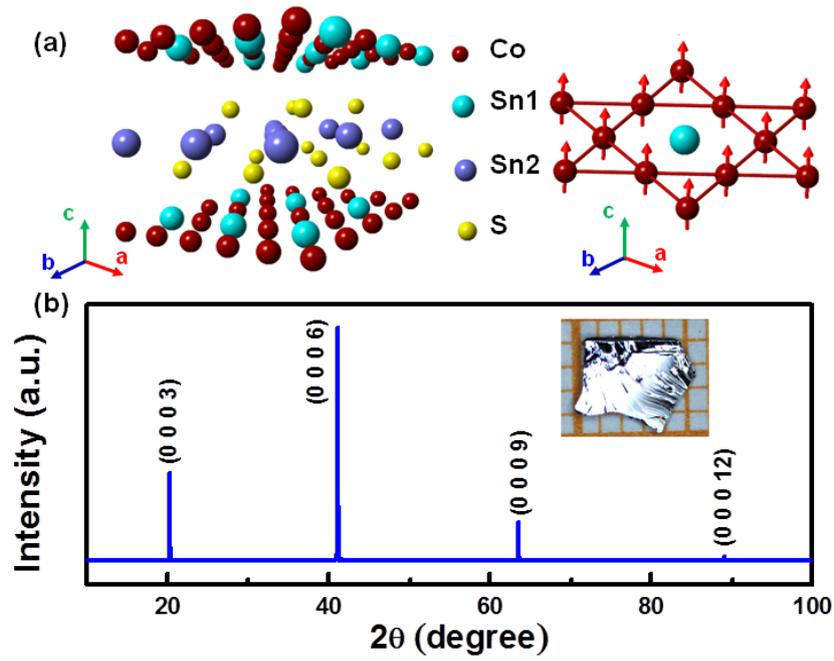

**Fig. 1. (a)** Crystal structure of $Co_3Sn_2S_2$ and kagome layer composed of Co atoms. **(b)** XRD pattern of $Co_3Sn_2S_2$ single crystal at room temperature. The inset shows a photograph of $Co_3Sn_2S_2$ single crystals.

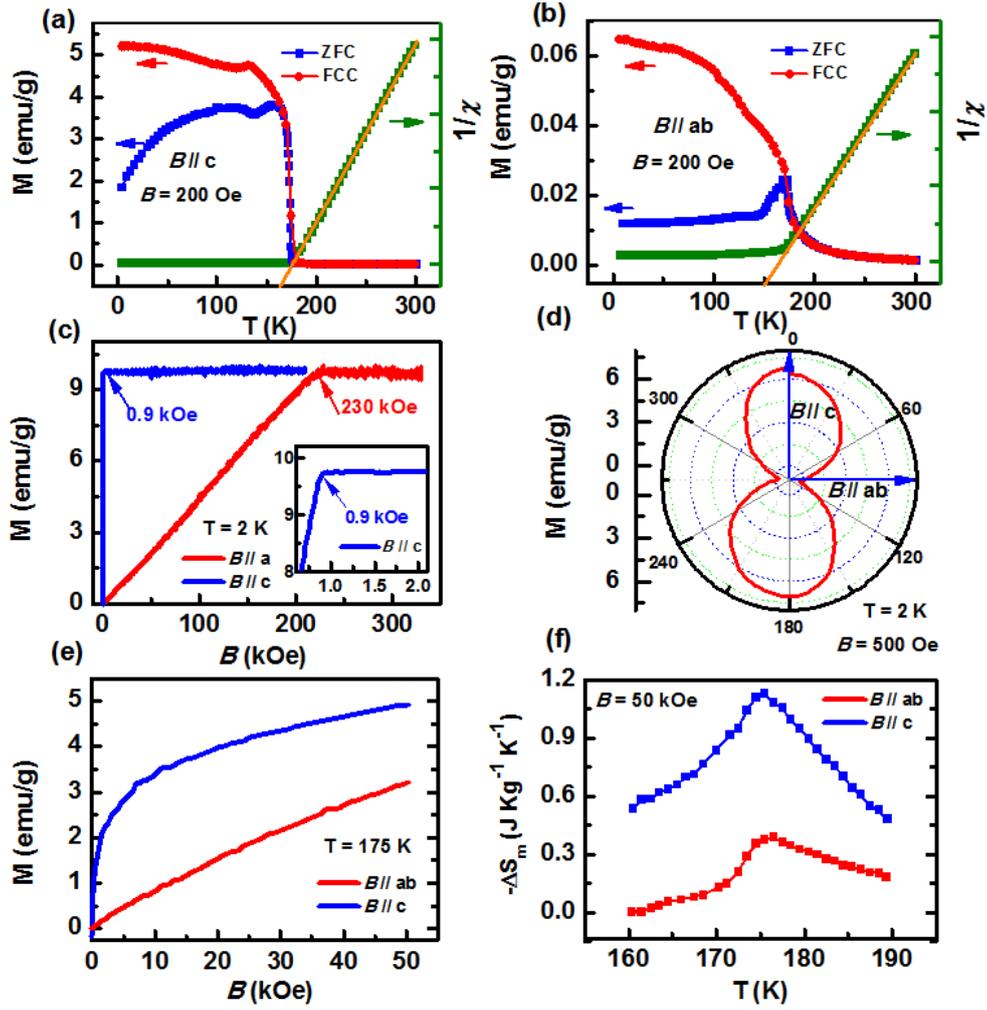

**Fig. 2. (a)** and **(b)** Temperature dependence of magnetization with ZFC and FCC measurements for $B // c$ and $B // ab$ at $B = 200$ Oe (left) and $1/\chi$ vs T behavior (right) for $B // c$ and $B // ab$, respectively. **(c)** Field dependence of magnetization at 2 K for $B // c$ and $B // ab$, respectively. The inset is the low field part for $B // c$. **(d)** Angular dependent magnetization from out-of-plane to in-plane at 2 K measured in $B = 500$ Oe. **(e)** Field dependence of magnetization at 175 K for $B // c$ and $B // ab$, respectively. **(d)** Temperature dependence of magnetic entropy change $-\Delta S_m$ at $B = 50$ kOe for $B // c$ and $B // ab$, respectively.

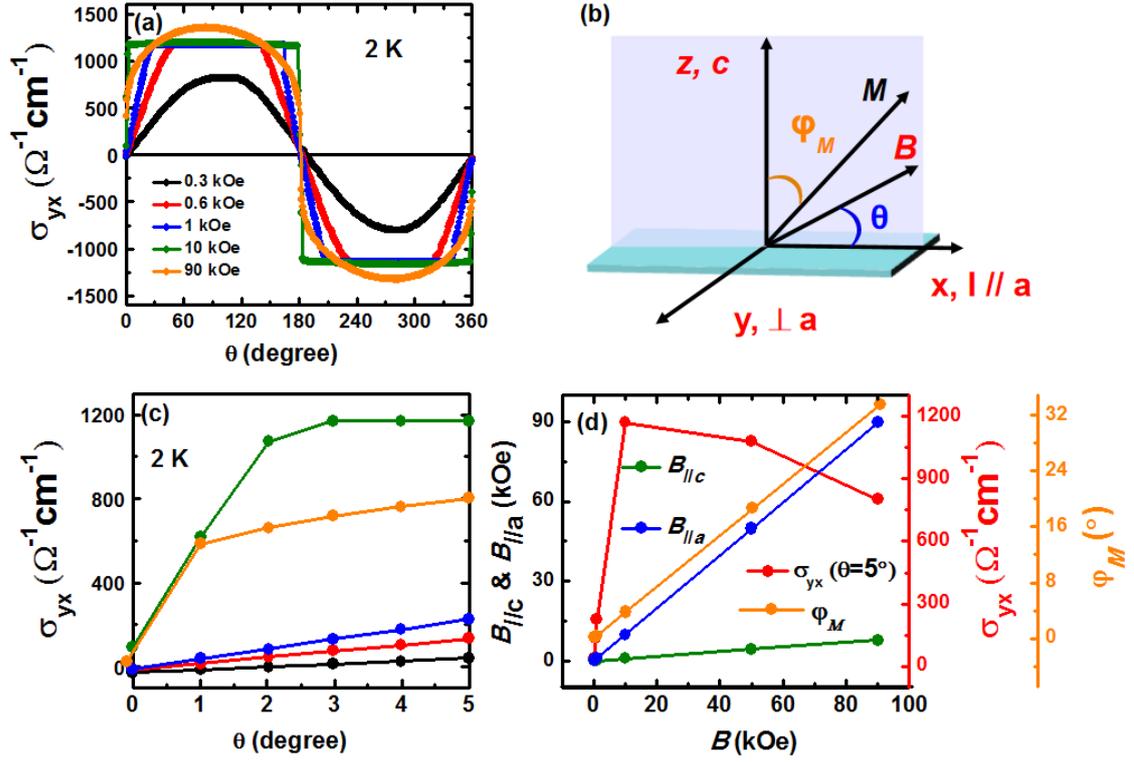

**Fig. 3. (a)** Angular (θ) dependence of Hall conductivity $\sigma_{yx}$ with different magnetic field rotating around *y*-axis in *ac*-plane at 2 K. Where θ is the angle between the magnetic field and *a* axis. **(b)** The configuration of current and magnetic field. **(c)** Hall conductivity $\sigma_{yx}$ with different magnetic field at θ = 0-5°. **(d)** Components of magnetic field in-plane ($B_a$) and out-of-plane ($B_c$), Hall conductivity $\sigma_{yx}$ and angle $\varphi_M$ between magnetic moment and *c*-axis at θ = 5°.

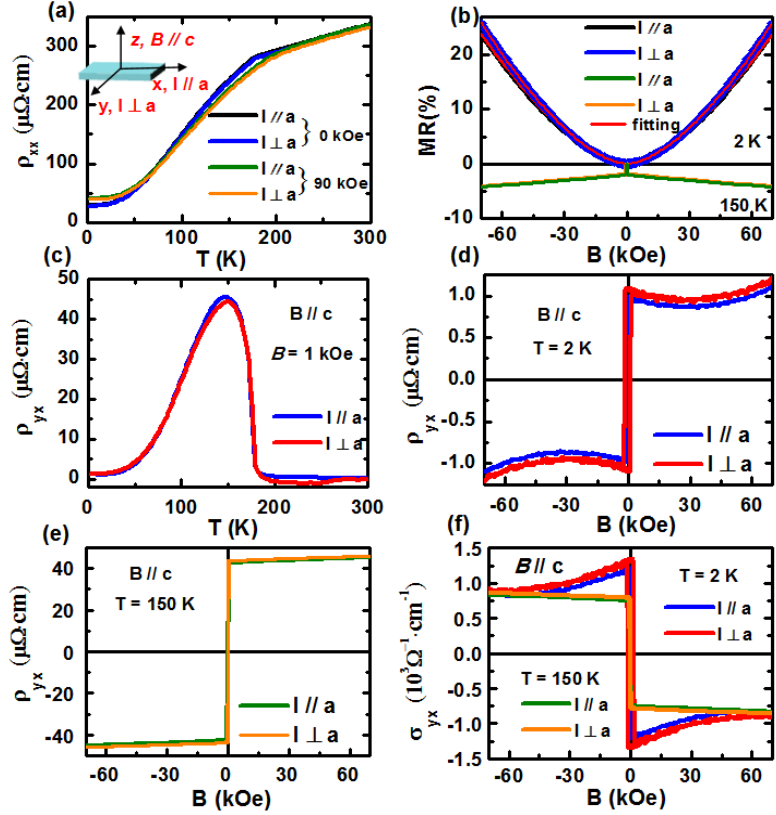

**Fig. 4 (a)** Temperature dependences of longitudinal resistivity $\rho_{xx}$ at zero field and 90 kOe for I // a and I ⊥ a, respectively. **(b)** Magnetoresistance as a function of magnetic field $B$ at 2 and 150 K for I // a and I ⊥ a with B // c, respectively. The fitting of positive MR at 2 K based on MR = $AB^{\alpha}$ for I // a and I ⊥ a. **(c)** Hall resistivities $\rho_{yx}$ as a function of temperature for I // a and I ⊥ a with B = 1 kOe, respectively. **(d)** and **(e)** Hall resistivities $\rho_{yx}$ as a function of magnetic field $B$ at 2 and 150 K for I // a and I ⊥ a with B // c, respectively. **(f)** Hall conductivities $\sigma_{yx}$ as a function of magnetic field $B$ at 2 and 150 K for I // a and I ⊥ a with B // c, respectively.

**Table 1.** Magnetocrystalline anisotropy coefficient ($K_u$) of some materials

| Systems | $Fe_3Sn_2$ | $Fe_3GeTe_2$ | $CrI_3$ | $CrBr_3$ | $Co_3Sn_2S_2$ |
|---|---|---|---|---|---|
| $K_u$ ($10^5$ J m$^{-3}$) | 3.3 | 0.69 | 0.86 | 3.0 | 8.3 |

**Table 2.** Anomalous Hall resistivity ($\sigma_{yx}^A$), carrier density of hole ($n_h$), carrier density of electron ($n_e$), mobility of holes ($\mu_h$), mobility of electron ($\mu_e$) of 2 K and maximal anomalous Hall angle ($\sigma_{yx}^A/\sigma_{xx}$) of $Co_3Sn_2S_2$ for I // $a$ and I $\perp$ $a$.

| I | $\sigma_{yx}^A$ $\Omega^{-1}$ cm$^{-1}$ | $n_h$ $10^{20}$ cm$^{-3}$ | $n_e$ $10^{20}$ cm$^{-3}$ | $\mu_h$ cm$^2$ V$^{-1}$ s$^{-1}$ | $\mu_e$ cm$^2$ V$^{-1}$ s$^{-1}$ | $\sigma_{yx}^A/\sigma_{xx}$ |
|---|---|---|---|---|---|---|
| I // $a$ | 1230 | 1.31 | 1.14 | 845 | 943 | 18.5% |
| I $\perp$ $a$ | 1300 | 1.24 | 1.2 | 870 | 918 | 19% |